\documentclass[sigconf]{acmart}

\usepackage{algorithm}

\usepackage{graphicx} 
\usepackage{graphics}
\usepackage{xspace}
\usepackage{makecell}
\usepackage{colortbl}
\usepackage{enumitem}
\usepackage{pifont}
\usepackage{amsfonts}
\usepackage{float}
\usepackage{subfig}
\usepackage{fontawesome5}
\usepackage{multirow}
\usepackage{subfig}
\usepackage{textcomp}
\usepackage{stfloats}
\usepackage{url}
\usepackage{verbatim}

\newcommand{\appname}{{\sc Diffploit}\xspace}

\newcommand{\testrepairname}{{\sc TaRGET}\xspace}
\newcommand{\ideaname}{{\sc IDEA}\xspace}
\newcommand{\appnamebold}{{\sc \textbf{Diffploit}}\xspace}

\newcommand{\ideanamebold}{{\sc \textbf{IDEA}}\xspace}
\newcommand{\testrepairnamebold}{{\sc \textbf{TaRGET}}\xspace}

\definecolor{myyellow}{HTML}{FFF2CC}
\definecolor{myblue}{RGB}{255,255,255}

\definecolor{lightgray}{gray}{0.95}

\AtBeginDocument{%
  }

\copyrightyear{2026}
\acmYear{2026}
\setcopyright{cc}
\setcctype{by}
\acmConference[ICSE '26]{2026 IEEE/ACM 48th International Conference on Software Engineering}{April 12--18, 2026}{Rio de Janeiro, Brazil}
\acmBooktitle{2026 IEEE/ACM 48th International Conference on Software Engineering (ICSE '26), April 12--18, 2026, Rio de Janeiro, Brazil}
\acmPrice{}
\acmDOI{10.1145/3744916.3773205}
\acmISBN{979-8-4007-2025-3/26/04}

\author{Zirui Chen}
\orcid{0009-0004-6236-9150}
\affiliation{%
  \institution{The State Key Laboratory of Blockchain and Data Security, Zhejiang University}
   \city{Hangzhou}
  \country{China}
}
\email{chenzirui@zju.edu.cn}

\author{Zhipeng Xue}
\orcid{0000-0002-9060-4064}
\affiliation{%
  \institution{The State Key Laboratory of Blockchain and Data Security, Zhejiang University}
   \city{Hangzhou}
  \country{China}
}
\email{zhipengxue@zju.edu.cn}

\author{Jiayuan Zhou}
\orcid{0000-0002-5181-3146}
\affiliation{%
  \institution{Queen’s University}
   \city{Kingston}
  \country{Canada}
}
\email{jiayuan.zhou@queensu.ca}

\author{Xing Hu}
\orcid{0000-0003-0093-3292}
\authornote{Corresponding authors}
\affiliation{
  \institution{The State Key Laboratory of Blockchain and Data Security, Zhejiang University}
   \city{Hangzhou}
  \country{China}
}
\email{xinghu@zju.edu.cn}

\author{Xin Xia}
\orcid{0000-0002-6302-3256}
\authornotemark[1]
\affiliation{
  \institution{The State Key Laboratory of Blockchain and Data Security, Zhejiang University}
   \city{Hangzhou}
  \country{China}
}
\email{xin.xia@acm.org}

\author{Xiaohu Yang}
\orcid{0000-0003-4111-4189}
\affiliation{%
  \institution{The State Key Laboratory of Blockchain and Data Security, Zhejiang University}
   \city{Hangzhou}
  \country{China}
}
\email{yangxh@zju.edu.cn}

\begin{document}

\title[Diffploit: Facilitating Cross-Version Exploit Migration for Open Source Library Vulnerabilities]{Diffploit: Facilitating Cross-Version Exploit Migration for \\ Open Source Library Vulnerabilities}

\begin{abstract}

Exploits are commonly used to demonstrate the presence of library vulnerabilities and validate their impact across different versions. However, their direct application to alternative versions often fails due to breaking changes introduced during evolution. These failures stem from both changes in triggering conditions (e.g., API refactorings) and broken dynamic environments (e.g., build or runtime errors), which are challenging to interpret and adapt manually.
Existing techniques primarily focus on code-level trace alignment through fuzzing, which is both time-consuming and insufficient for handling environment-level failures. Moreover, they often fall short when dealing with complicated triggering condition changes across versions. 
To overcome this, we propose \appname, an iterative, diff-driven exploit migration method structured around two key modules: the Context Module and the Migration Module. The Context Module constructs contexts derived from analyzing behavioral discrepancies between the target and reference versions, which capture the failure symptom and its related diff hunks. Leveraging these contexts, the Migration Module guides an LLM-based adaptation through an iterative feedback loop, balancing exploration of diff candidates and gradual refinement to resolve reproduction failures effectively. We evaluate \appname on a large-scale dataset containing 102 Java CVEs and 689 version-migration tasks across 79 libraries. \appname successfully migrates 84.2\% exploits, outperforming the change-aware test repair tool \testrepairname by 52.0\% and the rule-based tool in \ideaname by 61.6\%.
Beyond technical effectiveness, \appname identifies 5 CVE reports with incorrect affected version ranges, three of which have been confirmed. We also discover 111 unreported versions in GitHub Advisory Database.

\end{abstract}

\ccsdesc[500]{Security and privacy~Software security engineering}

\keywords{Library Vulnerabilities, Exploit Migration, Affected Version}

\maketitle

\section{Introduction}

Open-source libraries serve as critical infrastructure in modern software development, allowing developers to avoid redundant reimplementation and accelerate the development process~\cite{Wu2024Vision,Markus1,Synopsys1,Kula1, Zhan2025React}. However, the widespread adoption of open-source libraries raises concerns about the risk posed by vulnerabilities in these libraries~\cite{Zhou2024Magneto, zirui2024exploiting,He2023Dependent, shuhan2025vul, Zhan2024PS3, Pan2024Assessment, Pan2023Type, Pan2022Reports,jiayuan2022fix,jiayuan2023cole}, as they can propagate from upstream libraries to downstream projects~\cite{Bavota1,kula2018developers,Li2023downstream}. Given that libraries often evolve rapidly and downstream projects may depend on a wide range of versions, it becomes essential to assess the impact of vulnerabilities across versions~\cite{Shi2023Web, Bao2022Vszz}. To this end, developers often employ publicly disclosed exploits to evaluate whether and how vulnerable behavior manifests in different versions~\cite{Zhang2024SymBisect}, such as identifying affected library versions~\cite{Dai2021Exploit,Jiang2023AEM} and assessing exploitability in downstream projects with various affected versions~\cite{Zhou2024Magneto,zirui2024exploiting, Hong2022Poc, gao2025exploit, Deng2025Chainfuzz}.

However, simply reusing the original exploit on other affected versions frequently fails to reproduce without adaptation~\cite{Dai2021Exploit,Zhang2024SymBisect}, since disclosed exploits are typically crafted for the version in which the vulnerability was originally reported and do not generalize across other affected versions. These failures often result from (1) broken dynamic environments~\cite{Javan2023library,Zhang2024SymBisect,xue2024selfpico} and (2) changes in the underlying triggering condition~\cite{Zhong2024API}. Manually understanding and resolving these issues is often time-consuming and requires substantial expertise, highlighting the need for automated exploit migration.
Existing studies have demonstrated the feasibility of migrating exploits across versions~\cite{ Dai2021Exploit, Jiang2023AEM}, typically by aligning execution traces via API matching.   Despite their success in handling minor API changes, these approaches often rely on fuzzing, which can be time-consuming, especially when multiple versions require migration.  
More importantly, they overlook two critical challenges that frequently arise in practice:

\ding{182} Broken Dynamic Environment:
Previous methods primarily address code-level variations, neglecting the incompatibilities introduced by environmental changes such as dependency upgrades or runtime configurations. For instance, the exploit for CVE-2020-5245 \cite{exploit5245vision} fails with a runtime error in version \textit{1.3.8} due to an updated library (\textit{javassist}), despite identical exploit code. Such environment-induced failures may surface across multiple phases, including the build process and runtime~\cite{Yu2025Craft}, and exhibit considerable variability across libraries, posing significant challenges to mitigation.

\begin{figure*}[htbp] 
  \centering	
  
  \includegraphics[width=0.93\linewidth]{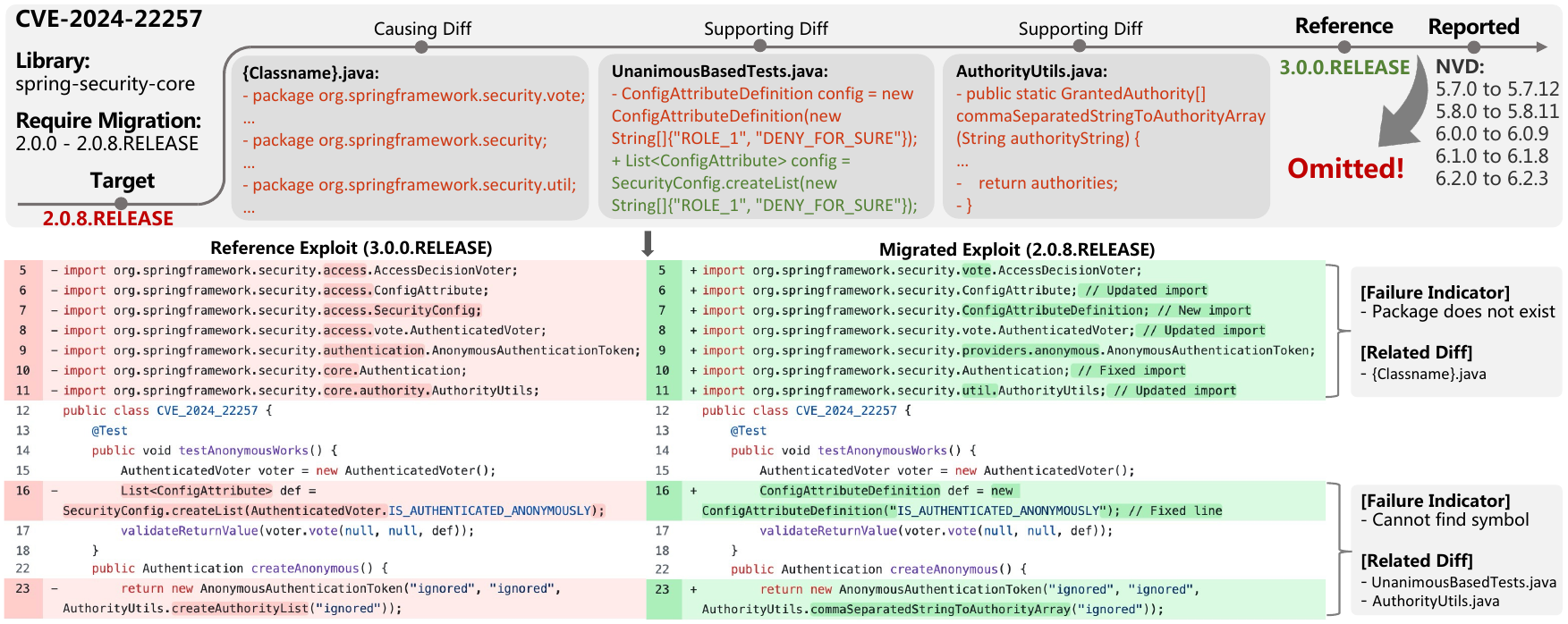}
  
  \caption{Migrating the exploit of CVE-2024-22257 from version 3.0.0.RELEASE to 2.0.8.RELEASE by \appnamebold.}
  \label{fig:example}
\end{figure*}

\ding{183} Complicated Triggering Condition Evolution:
When API interfaces undergo substantial changes, refactoring~\cite{wang2025refactoring}, or removal between versions, existing trace alignment methods, such as identifying function renaming or function merging/splitting~\cite{Dai2021Exploit}, struggle to identify suitable replacements. A representative example is CVE-2024-22257 ~\cite{NVD22257vision} illustrated in Figure ~\ref{fig:example}, where critical methods (\textit{createList} and \textit{createAuthorityList}) are absent in \textit{2.0.8.RELEASE}, requiring alternative APIs with significantly different signatures and semantics, which cannot be effectively addressed through existing API matching strategies. Incorrect or inadequate matching in these scenarios leads directly to ineffective migration efforts.

Addressing these challenges requires a comprehensive approach that considers both environmental adjustments and intelligent API adaptation mechanisms, beyond simple trace or naming alignment.
To effectively address the challenge of exploit migration across versions, we introduces an iterative, diff-driven LLM (Large Language Model) framework, \appname. The key idea is to iteratively adapt failed exploits using version-aware context and feedback. When a migration attempt fails, \appname compares the behavior of target version against a successful reference version to identify failure indicators. These indicators trigger the construction of a structured migration context by the \textbf{Context Module}: \textit{Causing Diffs} that likely led to the failure, and \textit{Supporting Diffs} that may help resolve it.
This context guides the \textbf{Migration Module} to adapt the exploit. A simulated annealing strategy is designed to explore and refine adaptations across multiple iterations. After each attempt, the updated exploit is re-executed, and new feedback is collected to guide the next cycle. Through this closed-loop process, \appname incrementally reduces discrepancies and achieves effective migration, even in the face of significant API or environment changes.

We evaluate \appname on a large-scale open source dataset of Java library vulnerability exploits~\cite{Wu2024Vision}, comprising 102 CVEs and their explicitly affected versions across 79 libraries. Among them, we identify 30 exploits that require migration across 689 versions, which is comparable to prior studies in the C/C++ domain~\cite{Dai2021Exploit}, involving 30 CVEs and 470 versions. Specifically, \appname successfully migrates 580 out of the 689 exploits to target versions in 23 CVEs, achieving a success rate of 84.2\%. Compared to the recent change-aware test repair research~\cite{Saboor2025repair}, \appname outperforms \testrepairname by 52.0\% and surpasses the rule-based approach in \ideaname by 61.6\%, highlighting the effectiveness of \appname in harnessing the capabilities of LLMs for exploit migration. We demonstrate the rationality of \appname's design through ablation studies, where it achieves a 46.1\% improvement over the base model. To validate the practicality and correctness of exploits migrated by \appname, we identify previously undisclosed vulnerable versions based on the migrated exploits on five CVE vulnerabilities and contact the CNAs   (CVE Numbering Authority)~\cite{cna} for confirmation. Among them, three cases are confirmed, with our submitted exploit links incorporated into the reference links. Meanwhile, we submit six pull requests to the GitHub Advisory Database to include a total of 111 missing versions detected by \appname, of which 82 versions are accepted. We further demonstrate that \appname effectively overcomes the previously mentioned challenges with a low cost.

This paper makes the following main contributions:

\begin{itemize}[leftmargin=*]

\item {We propose \appname, a novel discrepancy-driven approach that automatically migrates exploits from target versions to reference versions. Both the source code and dataset of \appname are available on our website~\cite{Replication}}

\item {We conduct an evaluation on 689 versions and demonstrate that \appname successfully migrates exploits for 580 of them, outperforming our baselines.}

\item {Our migrated exploits reveal five CVEs with inaccurate affected version ranges (three confirmed by CNAs) and uncover 111 affected versions missing from the GitHub Advisory Database.}

\end{itemize}

\section{Motivation}
In this section, we introduce the usage scenario and a motivating example to illustrate the challenges addressed by \appname.

\subsection{Usage Scenario}

In real-world vulnerability management, our method is designed to support two practical usage scenarios:

(1) Adapting ineffective exploits to confirmed affected versions.
Disclosed exploits are crucial for validating whether a specific project is affected by a known upstream vulnerability. However, these exploits are typically crafted for a particular version of a vulnerable library and often fail to function on other affected versions due to code or environment changes. For example, suppose a security advisory confirms that version 4.0.56 of the library \textit{netty-codec-http} is affected by CVE-2021-43797, but the public exploit only works on versions above 4.1.0.CR7~\cite{exploit5245vision}. Developers using version 4.0.56 are left without a working exploit to verify the vulnerability in their context. \appname addresses this gap by automatically adapting existing exploits to these confirmed yet incompatible versions, enabling practical verification in real-world environments.

(2) Discovering previously unreported affected versions.
Manually curated vulnerability reports often contain inaccuracies and omissions~\cite{Chaparro2017Version, Dong2019Version, Mu2018Report, Croft2023Report, Jo2021Report}. When a migrated exploit successfully triggers the vulnerability in a version not listed in public advisories, it suggests that the affected range may be broader than disclosed. Thus, \appname can assist in identifying missing affected versions and improving the completeness of vulnerability databases.
It is worth noting, however, that if \appname fails to migrate an exploit to a particular version, this does not necessarily mean the version is unaffected, and additional analysis is required~\cite{Jang2021Code}, such as introducing commit analysis~\cite{An2023Commit, Chen2025commit, xue2023acwrecommender}.

\subsection{Motivating Example}

Figure~\ref{fig:example} presents a motivating example from \textit{spring-security-core}, demonstrating how \appname facilitates exploit migration across versions. Through this process, we identify that versions prior to \textit{5.7.0.RELEASE} are affected by CVE-2024-22257, although they are not included in the CVE report. For versions above \textit{3.0.0.RELEASE}, a publicly disclosed exploit for CVE-2024-22257 is available to reproduce the vulnerability, which serves as the reference exploit. The reference exploit fails to execute directly on earlier versions, such as \textit{2.0.8.RELEASE}, because it depends on APIs introduced in later versions (e.g., line 16, line 23) and is affected by breaking changes introduced during updates (refactoring existing APIs to various packages)~\cite{Spring_security_commit}, which requires exploit migration. 

 Recent advances in artificial intelligence, especially LLMs, demonstrate strong capabilities in various tasks~\cite{grotov2024untanglingknotsleveragingllm, tyen2024llmsreasoningerrorscorrect, Yang2024AI, Li2024AI, Pan2024Patch, zou2024docbenchbenchmarkevaluatingllmbased, Saboor2025repair}, like understanding complex documentation~\cite{zou2024docbenchbenchmarkevaluatingllmbased} and resolving errors~\cite{grotov2024untanglingknotsleveragingllm, tyen2024llmsreasoningerrorscorrect}.
However, without sufficient contextual information, LLMs often struggle to update exploits using the appropriate APIs of the target versions.  Furthermore, in earlier versions, LLMs frequently fail to generate valid API calls due to the lack of corresponding usage patterns in their training data~\cite{spracklen2025packageyoucomprehensiveanalysis}. These limitations hinder the effectiveness of LLMs in generating reliable exploits tailored to target versions. We observe that providing diffs between the target version and the reference versions to LLMs enables them to infer the root causes of the failures and identify adaptation strategies, such as locating alternative APIs.

Based on this insight, we identify two types of diffs that facilitate exploit migration: \textit{Causing Diff} and \textit{Supporting Diff}. Failures in exploit reproduction, either due to modifications in triggering conditions or environmental breaks, are caused by changes introduced in the evolution of the library \textit{(causing diff)}. Meanwhile, libraries themselves often contain internal adaptations to these changes, which can provide LLMs with guidance for migration \textit{(supporting diff)}. Take CVE-2024-22257 as an example:

\begin{itemize}[leftmargin=*]
    \item {\textit{Causing Diff}}: The cause of the \textit{Package does not exist} error from line 5 to line 11 is the result of refactoring or modifications to the class file. Specifically, the corresponding hunks are those involving target package name changes, such as the removal of the package declaration line \textit{package org.springframework.security} in the \textit{ConfigAttribute} class. These changes provide LLMs with accurate package structure information of the target version, thereby reducing hallucinations such as generating outdated or non-existent import paths.
    
    \item {\textit{Supporting Diff}}: For the modification in line 16, a change in \textit{UnanimousBasedTests} reflects a response to the refactoring of \textit{ConfigAttributeDefinition} and guides the modification in resolving the \textit{Cannot find symbol} error for the \textit{createList} method. In addition, the diff in \textit{AuthorityUtils} provides explicit guidance for replacing the unavailable \textit{createAuthorityList} method at line 23 with an alternative API introduced in the target version. These diffs facilitate LLMs in addressing the evolution of triggering conditions by identifying a suitable replacement for deleted APIs.
\end{itemize}

Building on the identified diffs, we propose \appname, which iteratively leverages the diffs to facilitate exploit migration through a feedback loop. \appname takes nine steps to migrate the exploit for CVE-2024-22257, despite the significant challenge of identifying relevant information from the 2,555 diffs between the two versions.

\section{Proposed Approach}

In this section, we present our approach, \appname, detailing the overall workflow as well as the design of its two core components: the \textbf{Context Module} and the \textbf{Migration Module}.

\subsection{Overview}

Given a target version requiring exploit migration, our method iteratively repairs its behavioral discrepancies with reference exploits through a self-adaption process. The reference version is chosen from the set of reproduced versions to provide contextual guidance.

The overall process follows a discrepancy-driven loop. At each iteration, we compare the execution outputs of the target version and the reference version to identify reproduction failures. Each detected failure indicator in the target version triggers the construction of a dedicated migration context via the \textbf{Context Module}. This context encapsulates the failure symptom, its corresponding diagnostic key, and the associated code-level differences that may either underlie or assist in resolving the failure.

For each detected failure, the corresponding migration context is passed to the \textbf{Migration Module}, which attempts to adapt the exploit with the context. The migration process is guided by a simulated annealing strategy that explores diffs within the context to identify those effective in resolving the failure. The search process continues until the failure is resolved or a termination criterion (e.g., iteration limit or temperature threshold) is met.

Upon resolving each failure indicator, the adapted exploit is re-executed on the target version to collect fresh output. These outputs are then compared again to extract updated failure indicators.  The process iterates through these indicators until all discrepancies are either successfully addressed or deemed irrecoverable under the current search configuration.

\begin{figure*}[htbp] 
  \centering	
  
  \includegraphics[width=0.98\linewidth]{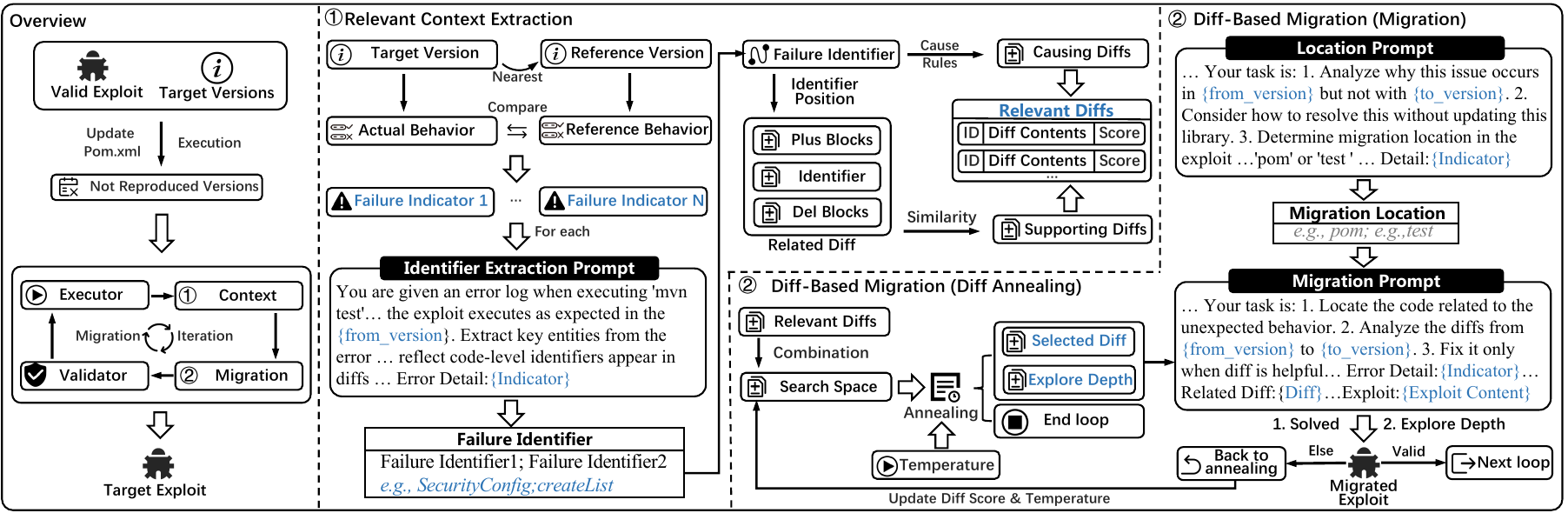}
  
  \caption{The overall framework and prompt details of \appnamebold.}
  \label{fig:overview}
\end{figure*}

\subsection{Context Module}

The Context Module extracts a set of \textit{migration contexts} from the target version $v_t$ by analyzing behavioral discrepancies with a reference version $v_r$. Each migration context corresponds to a specific failure unit, representing a reason for reproduction failure (triggering condition change or dynamic environment broken), and includes the related diffs that potentially cause or resolve the failure. These migration contexts provide structured and targeted information to guide the subsequent migration process.

\begin{definition}[Migration Context]
An \textit{migration context} is a structured representation that links a reproduction failure indicator $f_i$ observed in $v_t$ to its corresponding failure identifier $k_i$ , along with related diff hunks. Formally, it is defined as:

\[
\mathcal{C}_i = (f_i,k_i, \; D^{(i)}_{\text{cause}},\; D^{(i)}_{\text{support}})
\]
where:
\begin{itemize}[leftmargin=*]
    \item $D^{(i)}_{\text{cause}}$ are diff blocks related to the potential root cause of $f_i$;
    
    \item $D^{(i)}_{\text{support}}$ are diff blocks that provide additional assistance for migration, such as modified test cases or alternative functions.
\end{itemize}
\noindent{All diff blocks refer to Git-style diff hunks produced by \textit{git diff}.}
\end{definition}

\subsubsection{Failure Indicator Extraction} For each $v_t$ requiring exploit migration, we determine a corresponding  $v_r$ to serve as contextual guidance. The reference version is selected from a list of successfully reproduced versions $R = \{v_1, v_2, ..., v_n\}$ such that $v_r = \arg\min_{v \in R} \text{dist}(v, v_t)$, where $\text{dist}(\cdot)$ denotes a version distance metric. The list $R$ is constructed by executing the disclosed exploit across all historical versions from the Maven Central repository.

After identifying $v_r$, we execute the reference exploit against both $v_r$ and $v_t$, and collect respective outputs. We then compare the two outputs to detect behavioral discrepancies. Each discrepancy is abstracted into a \textit{failure indicator} $f_i$, representing a concrete symptom of the failure in $v_t$. Failure indicators encompass a variety of failure types, including build-time errors, runtime exceptions, and assertion failures. The set of extracted indicators $\{f_1, f_2, ..., f_n\}$ serves as the entry point for constructing migration contexts.

\subsubsection{Failure Identifier Extraction}

Given a failure indicator $f_i$, the goal of this module is to derive a corresponding set of failure identifiers $k_i$ that concisely capture the semantic essence of the failure to support downstream diff retrieval. While certain indicators, such as build-time errors, follow structured formats that can be parsed with simple rules, others such as runtime exceptions or unexpected execution outcomes are more variable in syntax and semantics, making them difficult to handle with pattern-based heuristics.

To address this, for each  $f_i$, we prompt the LLM to generate a corresponding set $k_i$  of identifiers, where each element in $k_i$ is a salient token selected to reflect the core failure content in a way that facilitates matching against diff content. These identifiers $k_i$ typically includes method names, exception types, or symbolic tokens that are likely to appear in the associated diffs, and will later be used to retrieve causing and supporting diffs.

\subsubsection{Causing Diff Extraction}

The goal of this module is to identify diff blocks that potentially the root cause of a given failure $f_i$. We leverage a set of handcrafted heuristic rules tailored to common failure types to extract relevant diffs. These rules allow us to efficiently associate symbolic error traces with corresponding code modifications without relying on complex program analysis.

Given a failure indicator $f_i$ and its corresponding identifier set $k_i$, we define a mapping function:
\[
D^{(i)}_{\text{cause}} = \textsc{Rule}(\text{type}(f_i), k_i)
\]
where $\text{type}(f_i)$ denotes the high-level category of the error (e.g., `AssertionMismatch', `MissingMethod', `RuntimeError'). We implement the following rules, which are designed to handle not only triggering condition changes and environment-level breakages but also other unexpected scenarios encountered during migration:

\begin{itemize}[leftmargin=*]

     \item \textbf{AssertionMismatch}: Collect diff blocks that affect method in $k_i$ calls directly invoked in the exploit, especially those on the call stack leading to assertion failure.

    \item \textbf{RuntimeError}: Identify diff blocks that modify methods, classes, or environment associated with the exception type or message tokens found in $k_i$. These symbols typically reflect failed API usage or behavior changes in dynamic execution.
   
    \item \textbf{MissingClass / MissingPackage}: Identify diffs under the file corresponding to the missing class or package in $k_i$.
    
    \item \textbf{MissingMethod}: Locate diff blocks that modify or remove the method signature referenced in $k_i$.
    
    \item \textbf{IncompatibleType / WrongReturn}: Retrieve diff blocks that affect method return types, parameter types, or generic usage near the method identified by $k_i$.

    \item \textbf{Other}: For all other failure types, we retrieve diff blocks that modify elements (e.g., classes, methods, or fields) whose names match any identifier in $k_i$. This fallback rule ensures general applicability when more specific heuristics are not available.

\end{itemize}

We apply a set of rules to determine whether a diff hunk satisfies these conditions (e.g., pattern \verb|\b\w[\w\s<>,]*\s+{cause}\s*\(| is used to detect \textit{MissingMethod} cases). 
Each diff hunk matched by these heuristics is then assigned a relevance score (10 points for the first five categories and 2 points for each $k_i$ in \textit{Other}) to guide the subsequent migration process.

\subsubsection{Supporting Diff Extraction}

This module aims to identify diff blocks introduced by the library in response to breaking changes leading to \( f_i \), which can assist the exploit migration process. We observe that breaking changes, such as API deletions or refactorings that cause exploit failures, also affect internal API usages within the library. To maintain functionality and compatibility, the library often introduces corresponding code updates, including modifications to internal method calls, additions of wrapper functions, or adjustments to test cases. These responsive changes serve as valuable supporting for guiding migration.

Such responses frequently appear as newly added diff blocks that are structurally similar to deleted or modified code within the same hunk. To detect such patterns, we first locate occurrences of tokens from the failure identifier set $k_i$ within the diff. Then, around each occurrence, we extract contiguous diff blocks composed of lines sharing the same diff prefix. Specifically, we collect:
\begin{itemize}[leftmargin=*]
    \item $\mathcal{B}_{+}$: blocks of contiguous added lines (prefixed with `$+$') near the positions where tokens in $k_i$ appear;
    \item $\mathcal{B}_{-}$: blocks of contiguous deleted lines (prefixed with `$-$') near the positions where tokens in $k_i$ appear;
\end{itemize}

We compute token-level similarity between each added block $b_+ \in \mathcal{B}_{+}$ and corresponding deleted block $b_- \in \mathcal{B}_{-}$ using the normalized length of their longest common subsequence (LCS):
\[
\text{sim}(b_+, b_-) = \frac{2 \times |\text{LCS}(b_+, b_-)|}{|b_+| + |b_-|}
\]
where \( |\text{LCS}(\cdot)| \) denotes the number of matched tokens in LCS, and \( |\cdot| \) denotes the token count of a block. If the similarity exceeds a threshold \( \tau \), the diff hunk containing the block pair is considered a candidate hunk and is selected as part of the support diff set $D^{(i)}_{\text{support}}$. We assign each selected hunk a score based on this similarity (ranging from 0-10), which is then used to guide the subsequent diff annealing process.

Finally, the migration context \(\mathcal{C}_i\) for failure \(f_i\) is generated by associating the failure indicator and its identifier \(k_i\) with the extracted causing diffs \(D^{(i)}_{\text{cause}}\) and supporting diffs \(D^{(i)}_{\text{support}}\), forming a structured representation to guide the migration process.

\subsection{Migration Module}

To support effective migration, the Context Module provides candidate diff hunks derived from the migration context $\mathcal{C}_i$. These candidates form the search space for the Migration Module and play a central role in navigating towards a successful exploit migration. Building on this set, the Migration Module guides the migration process by selecting diffs with our annealing strategy and presenting them along with $f_i$ to the LLM, which implicitly determines the appropriate migration locations within the exploit.

\subsubsection{Diff Annealing}

To expand the diversity and complexity of candidate diffs for migration, we design an annealing-based mechanism to explore a broader diff space with controlled randomness. Here, \emph{complexity} refers not only to the number of diffs applied simultaneously to a given failure indicator \( f_i \), but also to the number of repair rounds permitted when new failure indicators emerge during the migration process of \( f_i \). This setting allows the system to consider both wider combinations of patches and deeper chains of repair actions, thereby enriching the potential solution space.

This mechanism operates over three categories of diffs: high-scoring causing diffs \( D_{\text{cause}} \), supporting diffs \( D_{\text{support}} \), and a newly synthesized set of composite diffs \( D_{\text{combo}} \). We begin by constructing \( D_{\text{combo}} \) through pairwise combinations of top-ranked diffs from \( D_{\text{cause}} \cup D_{\text{support}} \). For each combination, the new diff inherits a composite score, computed as the average of its constituent diffs’ scores. Together, \( D_{\text{cause}} \cup D_{\text{support}} \cup D_{\text{combo}} \) define the search space \( \mathcal{D} \) for this annealing process.

To initiate the annealing, we define a temperature parameter \( T \), which governs both the selection probability of diffs and the exploration depth for resolving the current failure indicator \( f_i \). The selection probability of a diff \( d \in \mathcal{D} \) is proportional to its normalized score and temperature:
\[
P(d) \propto \exp\left(\frac{s(d)}{T}\right),
\]
where \( s(d) \) denotes the score of diff \( d \). Higher temperatures bias the selection toward higher-scoring diffs.

In addition, the \emph{exploration depth}---i.e., the number of retry attempts made when new failure indicators emerge during migration---is inversely correlated with the temperature. At higher temperatures, the system prefers quick evaluations of promising diffs, while deeper exploration is deferred to later stages when the temperature cools down. This design ensures that the system prioritizes efficient evaluation of high-scoring diffs in early stages.

During each iteration:
\begin{itemize}[leftmargin=*]
    \item A diff \( d \in \mathcal{D} \) is sampled and applied to perform a migration targeting \( f_i \) via LLM.
    \item If the migration fails, the score of \( d \) is penalized, the global temperature \( T \) is decreased, and the diff combination list \( D_{\text{combo}} \) is refreshed based on the updated scores.
    \item A new diff is then selected, and the process repeats.
\end{itemize}

This annealing-based strategy enables a controlled exploration of the diff space, balancing \emph{exploitation} of high-scoring diffs and \emph{exploration} of more diverse or complex diff combinations.

\subsubsection{Exploit Migration}

Given a failure indicator $f_i$ extracted from the execution outputs of the target version, the Exploit Migration module aims to adapt the original exploit such that the expected vulnerability behavior is restored. This process takes as input a selected related diff $d_i$, which is expected to resolve the failure based on prior analysis.

We first utilize the LLM to localize the relevant modification site within the exploit, conditioned on the failure indicator $f_i$, its associated diagnostic key $k_i$, and the selected diff $d_i$.  These elements are encoded into a structured prompt, as illustrated in our Figure ~\ref{fig:overview}, which guides the LLM in pinpointing the modification location.

Based on this localization, \appname then instructs the LLM to generate an adapted exploit $\hat{e}_i$ by modifying the original exploit according to the changes reflected in $d_i$ at the identified position.
$\hat{e}_i$ is then executed on the target version. Its runtime output is analyzed to determine whether $f_i$ has been resolved.

\subsubsection{Migration Validation}

Following the adaptation of the exploit targeting the specific failure indicator \( f_i \) in the preceding module, \appname re-run the exploit to collect the updated failure indicators in order to determine whether \( f_i \) has been resolved.

\begin{itemize}[leftmargin=*]
    \item If \( f_i \) still appears in the failure list, the migration attempt is considered unsuccessful. The process then returns to the annealing process to select the next diff candidate. This iterative search continues until \( f_i \) is resolved or the search terminates due to reaching the temperature limit or timeout.
    
    \item If \( f_i \) no longer appears in the failure list, indicating successful resolution, we analyze the updated failure indicators further:
    \begin{itemize}[leftmargin=*]
        \item If no failure indicators remain, the migration process proceeds to perform the final reproduction verification by checking whether the assertions in the exploit behave as expected and whether the observed behavior matches the expected reproduction behavior. If both conditions are satisfied, the migration process terminates successfully.

        \item If the remaining failure indicators are all previously known (i.e., identified prior to addressing \( f_i \)), we consider \( f_i \) to have been successfully resolved and proceed to handle the remaining failures accordingly.

        \item If new failure indicators emerge after attempting to resolve \( f_i \), we initiate a limited number of additional repair attempts to address these emergent failures. The number of such attempts, denoted as the \emph{exploration depth}, is determined based on the current temperature \( T \), with higher annealing temperatures yielding shallower exploration. If the number of repair attempts exceeds the exploration budget derived from \( T \), the current migration path is terminated, and a new diff candidate is selected to resolve \( f_i \).

    \end{itemize}
\end{itemize}

\section{Experimental Setup}

\textbf{Research Questions.} Our experimental evaluation aims to answer the following research questions:

\begin{itemize}[leftmargin=*]

\item {\textbf{RQ1 (Effectiveness):} To what extent can \appname effectively migrate exploits to target versions?}

\item {\textbf{RQ2 (Ablation Study):} How does each component within \appname contribute to the overall migration process?}

\item {\textbf{RQ3 (Practical Feasibility):} Dose \appname acceptable in real-world scenarios, considering the cost and quality of exploits?}

\end{itemize}

We address RQ1 to evaluate the effectiveness of our diff-based exploit migration method and its superiority over existing approaches. We address RQ2 to assess the rationality of key components in our method, including the causing diff extraction module, the supporting diff extraction module, and our proposed diff annealing algorithm. We address RQ3 to verify whether the exploits generated by \appname and the under-reported vulnerable versions it identifies are recognized and accepted.

\subsection{Dataset}

To minimize bias in the exploit collection process, we evaluate the performance of \appname using the largest publicly available Java exploit dataset~\cite{Wu2024Vision}. This dataset comprises 102 CVE vulnerabilities, each with a corresponding exploit targeting a specific version and a set of manually verified affected versions, making it well-suited for assessing the effectiveness of exploit migration approaches.

To identify versions requiring exploit migration, we execute exploits on versions labeled as affected and examine execution results. The dataset includes assertions designed to verify reproduction, which means versions failing to satisfy assertions are marked as requiring migration. We identify 988 versions meeting this criterion. Further analysis is performed to confirm the presence of vulnerabilities in these versions, resulting in a final set of 689 truly affected versions for 30 vulnerabilities. This reduction is mainly attributed to 176 versions in CVE-2023-51080 and 72 versions in CVE-2021-43795 that are incorrectly labeled as vulnerable before the introduction of the vulnerability~\cite{hutoolcommit, armeriacommit}.

\subsection{Baselines}

To the best of our knowledge, no prior work has explored exploit migration across Java library versions, though third-party libraries play a crucial role in the Java ecosystem~\cite{he2021migration}. Existing studies~\cite{Dai2021Exploit, Jiang2023AEM} rely on fuzzing frameworks such as AFLGo~\cite{aflgo}, which are difficult to adapt to Java. We include the following four baselines: \ding{182} \testrepairname ~\cite{Saboor2025repair} is a pre-trained language model-based approach for automated function-level test repair, which treats test repair as a language translation task and leverages context information extracted from the test breakage. Although it is not specifically designed for exploit migration, we include \testrepairname as a baseline due to its strong performance in repairing broken JUnit test cases, which are structurally similar to the exploits in our dataset. \ding{183} \ideaname ~\cite{idea} is the combination of Quick Fix and Auto-import features provided by IntelliJ IDEA. These features assist developers in resolving compilation issues caused by API refactoring, missing imports, or outdated method signatures, which can partially mitigate exploit failures caused by changes in triggering conditions. \ding{184} GPT-4o and \ding{185} DeepSeek-v3 are two of the most advanced proprietary LLMs, demonstrating strong capabilities in both code understanding and generation. In this study, we examine whether existing SOTA LLMs can generalize to the task without any task-specific fine-tuning.

\subsection{Migration Success Criteria}

To ensure an exploit triggers the same vulnerability after migration, we perform validation along two dimensions: assertion consistency and behavioral verification. Assertion consistency refers to whether the migrated exploit exhibits the same assertion failure behavior as the original, indicating consistency at the assertion level. Behavioral verification involves checking fine-grained behavioral indicators in the output logs to determine whether the expected vulnerability behavior is preserved after migration. 

An exploit is considered successfully migrated to the target version if and only if it triggers the expected assertion, and the location and manifestation of the assertion are consistent with those in the reference version.

\subsection{Implementation}

In our experimental setup, we deploy a Docker environment based on Ubuntu 20.04. Following the configuration by Wu et al.~\cite{Wu2024Vision} for exploit collection, we select Java 11 as the runtime environment, specifically version 18.9 (version: build 11+28). We execute exploits using the \textit{mvn test} command. For the selection of LLMs, we employ a high-performance closed-source model, GPT-4o (snapshot as of 2024-11-20), alongside an open-source model with strong performance on code tasks, DeepSeek-v3 (snapshot 0324), to conduct our experiments. To improve reproducibility, the base model of \appname is DeepSeek-v3. A time constraint of 5 minutes is applied to \appname and baselines for each version during experiments. Since all libraries in our dataset are hosted on GitHub/GitLab, we use \textit{git diff} to generate the diff files.

Regarding the baseline setup, we reproduce \testrepairname using the fine-tuned weight provided in the replication package. Despite our best efforts, we were unable to reproduce \testrepairname on CVE-2020-13956 and CVE-2023-51075, as it requires the construction of valid exploits tailored to these vulnerabilities. For experiments involving IntelliJ IDEA, we use version 2025.1.3. We apply IntelliJ IDEA Quick Fix and Auto-import features to modify the exploit and run tests using \textit{mvn test} after no further modifications are possible. When suggestions are abundant, including cases like multiple renaming suggestions, we evaluate the top five candidates.

\section{Experimental Evaluation}

\begin{table*}[htbp]
  \centering
  \caption{Performance of \appnamebold and Baseline Methods on Exploit Migration}
  \label{tab:result}
    \resizebox{0.95\linewidth}{!}{%
  \begin{tabular}{l c c cc c c c cc}
    \toprule
    \multirow{2}{*}{\textbf{Library}} & 
    \multirow{2}{*}{\textbf{CVE}} & 
    \multirow{2}{*}{\shortstack{\textbf{Reference}\\\textbf{Version}}} & 
    \multicolumn{2}{c}{\textbf{Affected Versions}} & 
    \multirow{2}{*}{\appnamebold} & 
    \multirow{2}{*}{\testrepairnamebold} & 
    \multirow{2}{*}{\ideanamebold} & 
    \multicolumn{2}{c}{\textbf{LLM}} \\
    
    \cmidrule(lr){4-5} \cmidrule(lr){9-10} 
    & & & \makecell[c]{\small\textbf{Total}} & \makecell[c]{\small\textbf{Need Mig.}} & & & &\makecell[c]{\small\textbf{GPT-4o}} & \makecell[c]{\small\textbf{DeepSeek-v3}} \\
    
    \midrule

    \multirow{1}{*}{\makecell[l]{commons-fileupload}} 
      & CVE-2016-1000031 & 1.1.1 & 8 & 3 & $\underline{3}$ & 0 & 0 &  3 & 3 \\

 \rowcolor{lightgray}
    \multirow{1}{*}{cxf-rt-rs-security-xml} 
      & CVE-2014-3584 & 2.6.10 & 33 & 14 & 0 & 0  & 0 & 0 & 0 \\

    \multirow{1}{*}{\makecell[l]{dolphinscheduler-api}} 
      & CVE-2022-34662 & 2.0.0 & 20 & 12 & 0 &  0 & 0 &  0 & 0 \\

\rowcolor{lightgray}
    \multirow{1}{*}{dolphinscheduler-common} 
      & CVE-2023-49250 & 2.0.0-alpha & 44 &12   & 0 &  0 & 0& 0 & 0\\

    \multirow{1}{*}{dropwizard-validation} 
      & CVE-2020-5245 & 2.0.1 & 40 & 16 & $\underline{16}$ &  0 & 0  & 0 & 0\\

 \rowcolor{lightgray}
    \multirow{1}{*}{hibernate-validator} 
      & CVE-2019-10219 & 6.0.5.Final & 31 & 12 & $\underline{12}$ & 0 & 0 & 1 & 12\\

    \multirow{1}{*}{httpclient} 
      & CVE-2020-13956 & 4.5.3 & 40 & 6 & 0 & -- & 0 & 0 & 0 \\
    
\rowcolor{lightgray}
    \multirow{1}{*}{hutool-core} 
      & CVE-2023-51075 & 5.7.18 & 38 & 1 & 0 & -- & 0 & 0 & 0 \\

     \multirow{1}{*}{jackson-databind} 
      & CVE-2022-42004 & 2.0.0-RC1 & 24 & 5 & 2 & 0 & 0 & 0 & 0\\

\rowcolor{lightgray}
      \multirow{1}{*}{jackson-dataformat-xml} 
      & CVE-2016-7051 & 2.7.7 & 75 & 69 & $\underline{69}$ & 0 & 0& 0& 0\\

    \multirow{1}{*}{junrar} 
      & CVE-2022-23596 & 6.0.0 & 15 & 8 & $\underline{8}$ & 0 & 0 & 5 & 8\\

    \rowcolor{lightgray}
    \multirow{1}{*}{kernel} 
      & CVE-2022-24197 & 7.2.0 & 24 & 5 & 0 & 0 & 0 & 0 & 5 \\

    \multirow{3}{*}{netty-codec-http} 
      & CVE-2021-43797 & 4.1.49.Final & 157 & 86 & $\underline{86}$ & 80 & 0 & 80 & 0\\
      & CVE-2019-20444 & 4.1.43.Final & 131 & 71 & $\underline{71}$ & 65 &  65 & 1 & 65\\
      & CVE-2019-16869 & 4.1.0.Beta1 & 131 & 71 & $\underline{71}$ & 65 & 65 & 1 &65\\

   \rowcolor{lightgray}
    \multirow{1}{*}{netty} 
      & CVE-2015-2156 & 3.3.0.Final & 65 & 45 & 0 & 0 & 0 & 0 & 0\\

    \multirow{1}{*}{para-core} 
      & CVE-2022-1848 & 1.42.2 & 102 & 84 & $\underline{84}$ & 0 & 21 & 84 & 23 \\

    \rowcolor{lightgray}
    \multirow{1}{*}{plexus-utils} 
      & CVE-2017-1000487 & 1.4.2 & 50 & 2 & $\underline{2}$ & 2 & 0 & 2 & 2\\

    \multirow{1}{*}{postgresql} 
      & CVE-2024-1597 & 9.4.1212 & 179 & 49 & 48 & 0 & 0 & 0 & 0  \\

    \rowcolor{lightgray}
    \multirow{1}{*}{protocols-imap} 
      & CVE-2021-40111 & 3.5.0 & 11 & 9 & $\underline{9}$* & 0 & 0 & 0 & 0\\

    \multirow{1}{*}{socket.io-client} 
      & CVE-2022-25867 & 1.0.0 & 12 & 9 & $\underline{9}$ & 0 & 0 & 0 & 0\\

    \rowcolor{lightgray}
    \multirow{1}{*}{spring-amqp} 
      & CVE-2017-8045 & 2.1.0.RELEASE & 49 & 2 & $\underline{2}$ & 0 & 0 & 2 & 2\\

    \multirow{1}{*}{spring-actuator-logview} 
      & CVE-2021-21234 & 0.2.9 & 14 & 10 & $\underline{10}$ & 10 &  5 & 5 & 10\\

    \rowcolor{lightgray}
    \multirow{1}{*}{spring-context} 
      & CVE-2022-22968 & 4.2.9.RELEASE & 194 & 19 & $\underline{19}$ & 0 & 0 & 19 & 19\\

    \multirow{2}{*}{spring-security-core} 
      & CVE-2019-11272 & 3.0.0.RELEASE & 61 &  9 & $\underline{9}$ & 0 & 0 & 0 & 0\\
      & CVE-2024-22257 & 5.7.11 & 210 & 9 &  $\underline{9}$ & 0 & 0 & 0 & 9 \\

    \rowcolor{lightgray}
    \multirow{1}{*}{spring-webmvc} 
      & CVE-2014-3625 & 3.0.4.RELEASE & 30 & 1 & $\underline{1}$ & 0 & 0 & 1 & 1\\

    \multirow{2}{*}{spring-web} 
      & CVE-2013-6430 & 1.1.1 & 53 & 4 & \underline{4} & 0 & 0 & 0 & 0\\
      & CVE-2020-5421 & 5.2.8.RELEASE & 124 & 39 & $\underline{39}$ & 0 & 0 & 39 & 39 \\

    \rowcolor{lightgray}
    \multirow{1}{*}{wicket-core} 
      & CVE-2013-2055 & 1.5.10 & 33 & 7 & $\underline{7}$ & 0 & 0 & 0 & 0\\

    \midrule

    \multirow{1}{*}{SUM} 
      & 30 CVEs & -- & 1,998 & 689 & 580 (84.2\%) & 222 (32.2\%) & 156 (22.6\%) & 243 (35.2\%) & 263 (38.1\%)\\
    \bottomrule
  \end{tabular}
  }
  \vspace{2pt}
  \begin{minipage}{0.95\linewidth}
    \footnotesize *The exploit is a flaky test.
  \end{minipage}
\end{table*}

We evaluate the performance of \appname from three perspectives. First, we assess its effectiveness and compare it with baseline methods using the largest dataset of Java library vulnerability exploits, and analyze its strengths and limitations. Second, we conduct an ablation study to demonstrate the contributions of the diff and migration modules to the overall performance. Finally, we analyze the practical feasibility of \appname based on cost and responses from CNAs and open-source maintainers.

\subsection{Effectiveness}

\subsubsection{Performance}
We evaluate the effectiveness of \appname on a dataset containing 689 version pairs that require exploit migration. \appname successfully migrates 580 of them, achieving an overall success rate of 84.2\%. In the context of 30 representative CVEs, \appname successfully performs exploit migration for 23 cases, covering a diverse range of vulnerability types and affected libraries. This demonstrates the generality of \appname across both vulnerability classes and dependency ecosystems. Notably, for 20 CVEs, all associated versions requiring migration are successfully repaired, highlighting the potential of \appname to serve as a reliable component in automated vulnerability validation pipelines. However, it is important to emphasize that a failed migration does not necessarily imply the absence of a vulnerability.

Compared to the baseline method \testrepairname, \appname achieves a relative improvement of 51.0\%, demonstrating its robustness in handling critical challenges in exploit migration, such as adapting to non-function-level edits like import adjustments and build configuration updates. When compared with \ideaname that combines IntelliJ IDEA’s Quick Fix and Auto Import features, \appname outperforms significantly in 61.6\%. This result highlights its ability to address migration scenarios that exceed the capabilities of predefined rules. We also compare \appname against direct application of LLMs. GPT-4o and DeepSeek-V3 can migrate 243 (35.2\%) and 263 (38.1\%) cases respectively, which, while competitive in some straightforward scenarios, underperform due to lack of migration context. In contrast, \appname leverages migration-specific contextual information, enabling it to maintain a higher success rate especially in complex migration scenarios.

\subsubsection{Strength}

Compared to the baseline method \testrepairname, \appname demonstrates superior adaptability in handling test changes that are not confined to the function level. While \testrepairname focuses on identifying faulty functions, it relies on developers to manually apply legitimate edits when changes occur outside function bodies, such as modifications in import statements or resolving runtime environment broken in pom.xml. As a result, it fails to repair test cases in 16 CVEs where such non-functional changes are essential.

\appname is also capable of addressing scenarios where an exploit test should fail but instead passes silently in the target version. These cases, where the presence of a vulnerability is masked by a superficially successful test, are often overlooked by \testrepairname. By leveraging rich migration context and historical diffs, \appname can detect and adapt such misleading test cases, ensuring they remain effective indicators of vulnerabilities.

Additionally, \appname leverages a structured migration context that captures both the causing and supporting diffs of a test case. This design enables it to accommodate a broader range of code modifications related to exploit migration. This advantage becomes more prominent in complex exploit migration tasks, which require understanding subtle semantic changes. In our Discussion section, we further analyze the edit distance before and after migration, highlighting \appname's ability to apply complex adaptations.

\subsubsection{Limitations}
\label{sec:limitations}
While \appname demonstrates strong performance in migrating exploits across versions, it has several limitations. First, it struggles to handle cases where the vulnerability manifests differently across versions. For instance, in CVE-2020-13956 and CVE-2023-51075, the exploits trigger exceptions such as \textit{NumberFormatException} and \textit{IndexOutOfBoundsException}, rather than the originally expected assertion failures. Although such runtime exceptions still indicate security-relevant behavior, \appname currently treats them as failures, as it enforces a fixed assertion-based validation strategy. Similarly, in CVE-2023-49250, the exploit leads to a malicious server connection.

\begin{figure}[htbp] 
  \centering	
  
  \includegraphics[width=1\linewidth]{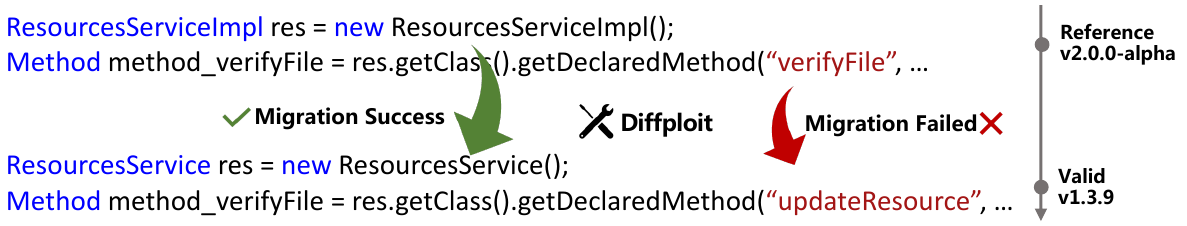}
  
  \caption{A failure case of \appnamebold for CVE-2022-34662.}
  \label{fig:failure}
\end{figure}

Second, \appname encounters difficulties when migrating exploits that depend on version-specific APIs, especially when the target version lacks both a structurally similar counterpart and supporting information. As illustrated in Figure \ref{fig:failure}, in CVE-2022-34662, the original exploit in a higher version invokes \textit{ResourcesServiceImpl.verifyFile}, while the lower version provides a semantically related method \textit{updateResource} under a different class, \textit{ResourcesService}. \appname failed to migrate the method due to missing contextual alignment. This example shows that while \appname can partially adapt such cases, external API knowledge is required.

\subsection{Ablation Study}

Our ablation study aims to achieve two goals: (1) to demonstrate that each component in our design contributes to higher exploit reproduction success, and (2) to show that our design helps reduce the reproduction cost in terms of step count. We construct three ablated variants of \appname to evaluate the contribution of each component:  
(a) \textbf{\appnamebold-Causing}, which disables the extraction of causing diffs;  
(b) \textbf{\appnamebold-Supporting}, which disables the extraction of supporting diffs; and  
(c) \textbf{\appnamebold-Annealing}, which removes the diff annealing process.  
To further evaluate the effectiveness of our diff combination strategy,  
we design an additional variant named (d) \textbf{\appnamebold-Combining},  
in which only diff scores are used to provide contextual information,  
without incorporating among diffs.
The performance of the base model without any diff information has already been evaluated in Table~\ref{tab:result}, so we do not include a separate variant for this setting.

We evaluate each variant using three metrics reported in Table~\ref{tab:ablation}.  
\textbf{Average Step} measures the average number of adaptation steps required to successfully migrate exploits, where failures are assigned a default value of 30 steps, derived from the estimated time to complete exploit execution and the response latency of the LLM within five minutes.
\textbf{Success Rate} reports the version number and percentage of exploits successfully migrated.  
\textbf{Overhead} quantifies the average step overhead relative to the origin \appname, offering a normalized view of the cost efficiency.

\begin{table}[htbp]
  \centering
  \caption{Ablation Study on \appnamebold.}
  \label{tab:ablation}
\resizebox{0.95\linewidth}{!}{%
  \begin{tabular}{cccc}
    \toprule
    \textbf{Method} &  \textbf{Average Step*} & \textbf{Success Rate} & \textbf{Overhead} \\
    \midrule
    \appname & 8.28 & 580/689 (84.2\%)  & - \\
    \appname-Causing & 12.71 & 470/689 (68.2\%)  & 277.15\%\\
    \appname-Supporting & 11.15 & 489/689 (71.0\%) & 159.09\%\\
    \appname-Annealing & 15.02  & 462/689 (67.1\%) & 444.69\%\\
    \appname-Combining & 13.35 & 505/689 (73.3\%) & 400.92\%\\
    \bottomrule
  \end{tabular}
    }
  \vspace{2pt}
  \begin{minipage}{0.95\linewidth}
    \footnotesize * Failures are assigned a default number of steps.
  \end{minipage}
\end{table}

We observe from Table~\ref{tab:ablation} that \appname achieves the highest success rate, outperforming the best ablated variant by over 10.9\%. This confirms that each component in our design contributes to the overall success. Among the variants, although the success rates degrade when removing any single module, they all remain notably higher than the base model (evaluated separately in Table~\ref{tab:result}), indicating that the use of context-aware diffs and annealing strategies brings substantial benefits. In terms of cost, \appname not only achieves the highest success rate but also requires the fewest steps on average (8.28), demonstrating that each component of \appname contributes effectively to the overall performance. Although all three variants yield similar success rates, they incur different levels of reproduction cost, with the variant without \textit{Annealing} performing the worst (15.02 steps on average, 444.69\% overhead relative to the origin \appname). This highlights the importance of the annealing process in filtering and prioritizing diffs that are most helpful to migration. Although \appname-Combining achieves the highest success rate among all variants, it requires significantly more steps than \appname, 
indicating that our combination strategy substantially improves the efficiency of discovering valid exploits.

\subsection{Practical Feasibility}

\subsubsection{Acceptability} 
We aim to introduce an objective assessment of \appname’s practical feasibility by assessing whether migrated exploits are accepted by real-world vulnerability management processes. Some adapted exploits target early branches that are no longer maintained, making it impossible to evaluate the acceptance of \appname by library maintainers. To overcome this, we evaluate the acceptance of \appname by submitting the previously undocumented affected versions supported by our migrated exploits. We identify and report previously undocumented affected versions to CVE, the most authoritative vulnerability repository, and the GitHub Advisory Database, which facilitates communication through pull request-based submissions. Our migrated exploits serve as supporting evidence in these submissions. We assess the real-world acceptance of \appname based on the responses from CNAs and open-source reviewers.

\appname successfully migrates exploits to 580 affected versions. We further investigate whether these versions are explicitly documented in the NVD descriptions and the affected product listings in the GitHub Advisory Database. As illustrated in Table~\ref{tab:realworld}, our migrated exploits uncover five NVD entries with missing or unclear specifications of affected version ranges, as well as 111 affected versions omitted from the GitHub Advisory Database.  We contact the corresponding CNAs via email, and three of them has incorporated our migrated exploit into the reference links of the CVE report, while the remaining CNAs have not responded by the time of submission.
We also submit six pull requests to the GitHub Advisory Database, which result in updates to 57 affected versions. The remaining submissions are not merged due to GitHub's limited capacity to validate exploits at scale~\cite{pr}.
The addition of our exploits to the CVE reference links and the update of affected versions identified by migrated exploits demonstrate our practical feasibility.

\begin{table}[htbp]
  \centering
  \caption{Real World Response of \appnamebold.}
  \label{tab:realworld}
    \resizebox{0.95\linewidth}{!}{%
  \begin{tabular}{cccc}
    \toprule
      \textbf{CNA} &\textbf{CVE} & \textbf{CNA Response} & \textbf{GitHub Missing}\\
    \midrule
     Mitre & CVE-2019-20444 & Confirmed & 14 (12 Accepted) \\
     Mitre & CVE-2019-16869 & Confirmed & 2 (0 Accepted) \\
     GitHub, Inc. & CVE-2021-43797 & Under Review & 13 (Under Review) \\
     GitHub, Inc. & CVE-2020-5245 & - & 19 (7 Accepted) \\
     VMWare &  CVE-2024-22257 & Under Review & - \\
      Red Hat, Inc. & CVE-2013-6430 & - & 38 (38 Accepted) \\
      Red Hat, Inc. & CVE-2019-10219 & Confirmed & 25 (25 Accepted) \\
      \midrule
        SUM & 7 CVEs & 3 Confirmed & 111 (82 Accepted) \\
    \bottomrule
  \end{tabular}
    }
\end{table}

\appname contributed to improving the quality of the CVE reports by identifying additional affected versions and providing a working exploit. As confirmed by Red Hat: \textit{``Thank you again for helping us improve the CVE records. The changes for version information were made, and the references you sent added.''}

\subsubsection{Cost} We estimate the financial cost of using \appname to migrate an exploit with base model DeepSeek-V3. On average, each successfully migrated exploit consumes 7,029 input tokens and 831 output tokens. According to the pricing details provided by DeepSeek as of July 2025, this corresponds to an average cost of \$0.0014 per successful migration. Similarly, for failed migrations, the average token consumption is 12,977 input tokens and 997 output tokens, resulting in an average cost of \$0.0020 per failed attempt. These results indicate that the financial cost of using \appname remains low and practical for real-world deployment.

\subsubsection{Data Leakage}

To evaluate the performance of \appname on unseen vulnerabilities, we conduct an experiment using exploits that are disclosed after the model cutoff date. Our base model used for all experiments is \textit{Deepseek-V3-0324}, thus we select five CVEs that are published after March 2025.

As illustrated in Table~\ref{tab:leakage-cves}, \appname successfully migrates exploits for all affected versions of four unseen CVEs, with only two versions of CVE-2025-53864 failing to produce valid exploits. These results demonstrate the robustness of our approach when handling vulnerabilities that the underlying LLM has never seen before.

\begin{table}[htbp]
  \centering
  \small
  \caption{Evaluation on Vulnerabilities Disclosed Post Cutoff.}
  \label{tab:leakage-cves}
  \resizebox{\linewidth}{!}{%
  \begin{tabular}{ccccc}
    \toprule
    \multirow{2}{*}{\textbf{CVE-ID}} & \multirow{2}{*}{\makecell{\textbf{Published}\\\textbf{Date}}}  & \multicolumn{2}{c}{\textbf{Affected Versions}} & \multirow{2}{*}{\textbf{Success}} \\\cmidrule(lr){3-4} & & \textbf{Total} & \textbf{Need Mig.} \\
    \midrule
    CVE-2025-59340 & 2025-09-17 & 92  & 19  & 19  \\
    CVE-2025-58056 & 2025-09-03 & 227 & 71  & 71  \\
    CVE-2025-53864 & 2025-07-10 & 268 & 192 & 190 \\
    CVE-2025-52999 & 2025-06-25 & 183 & 28  & 28  \\
    CVE-2025-48976 & 2025-06-16 & 12  & 2   & 2   \\
    \bottomrule
  \end{tabular}
  }
\end{table}

\section{Discussion}

This section first examines the extent to which \appname addresses the identified challenges. We then discuss potential threats to validity and how they might affect the interpretation of our results.

\subsection{Challenge Resolution}

We further discuss how \appname addresses the challenges we proposed. We examine the effectiveness of \appname in addressing two primary challenges: \ding{182} failures caused by environment broken, and \ding{183} the complexity of exploit migration process.

\ding{182} \textbf{Broken Dynamic Environments.} To understand whether \appname addresses failures caused by broken dynamic environments, we examined the modification locations in migrated exploits. Among the 23 exploits that are successfully migrated, seven of them do not require modifications to the exploit code. These migrations enabled \appname to successfully reproduce 128 previously failing versions, suggesting that many failures of origin exploits are not rooted in the exploit logic but rather caused by external or environmental factors. We further categorized the types of failures that \appname handled without modifying the exploit code. These include (1) incompatibilities caused by other dependencies required by the exploit, such as mismatched versions of \textit{spring-web} and \textit{spring-test}; (2) build-time issues caused by blocked or obsolete repositories in the Maven configuration, such as  \textit{repo.spring.io}; (3) runtime failures due to Java platform evolution (e.g., \textit{JAXB} in Java 11); and (4) dependency resolution failures related to outdated or snapshot artifacts (e.g., \textit{commons-io}, \textit{javax.faces:jsf-api}), or conflicts in third-party libraries (e.g., \textit{Javassist}). These results demonstrate that \appname effectively addresses failures stemming from environmental issues by reconstructing the necessary build and runtime context in the pom file.

\ding{183} \textbf{Complicated Migration.} We assess the migration complexity of \appname using the Average Edit Distance (AED), a widely adopted metric for migration tasks. AED is defined as the average token-level Levenshtein distance computed over a set of exploit pairs. Specifically, we calculate the AED between exploits before and after migration to evaluate the overall complexity of exploits migrated by \appname. Additionally, to estimate the manual effort required when relying solely on \ideaname for migration, we compute the AED between the exploits modified using \ideaname alone and the corresponding valid exploits produced by \appname. This comparison quantifies the reduction in manual effort achieved by \appname.

On the 22 CVEs where exploit migration is successful by \appname, the average edit distance between the migrated exploits and the original exploits is 176.00, indicating that the required modifications are non-trivial and reflect substantial structural or semantic changes. In contrast, \ideaname is able to provide useful suggestions for only 7 of these cases. Even after applying its suggestions, the resulting exploits still require an average edit distance of 59.69 to match valid exploits produced by \appname. Based on these observations, we estimate that \appname reduces manual effort by approximately 80\%, as measured by the total edit distance required for the exploit migration compared to \ideaname. This highlights \appname's ability to both handle complex migration scenarios and significantly alleviate the manual burden associated with exploit migration.

\subsection{Qualitative Analysis}

As discussed in Section~\ref{sec:limitations}, we have analyzed the limitations of our approach. 
To complement that discussion, we further conduct a qualitative analysis of failure cases, 
which can be broadly categorized into three root causes. Understanding these categories helps explain the remaining gaps and guide future improvement.

\textbf{Assertion mismatch.} Diffploit enforces the original assertions during migration. Some vulnerabilities exhibit different behaviors on different versions, leading to the original exploit assertion being ineffective on the target versions. Table~\ref{tab:qual-assertion} lists representative examples where the observed behavior diverged from the origin.

\begin{table}[h]
  \centering
  \small
  \caption{Examples of Assertion Mismatch.}
  \label{tab:qual-assertion}
  \resizebox{0.95\linewidth}{!}{%
  \begin{tabular}{ccc}
    \toprule
    \textbf{CVE} & \textbf{Expected} & \textbf{Actual} \\
    \midrule
    CVE-2020-13956 & Wrong Return Value        & NumberFormatException     \\
    CVE-2023-51075 & Timeout                   & IndexOutOfBoundsException \\
    CVE-2023-49250 & Console Output            & Remote Server Connection  \\
    CVE-2022-42004 & StackOverflowError        & Unexpected Exception      \\
    CVE-2024-1597  & Wrong Return Value        & VM crash                  \\
    \bottomrule
  \end{tabular}
  }
\end{table}

\textbf{Domain-specific inputs.} Some exploits require highly specialized, format-specific payloads that cannot be synthesized reliably from code diffs or shallow context. Examples include crafted PDFs (e.g., CVE-2022-24197) and structured tokens or cookie names (e.g., CVE-2014-3584, CVE-2015-2156). These artifacts often need precise byte-level or format-aware modification, which is outside the current capabilities of \appname.

\textbf{Large API gaps.} When the target version introduces substantial structural changes (multiple co-occurring edits, API renames, or replaced components), the search and edit synthesis of \appname may only partially complete the necessary transformations. For example, CVE-2022-34662 required replacing a service class and finding an appropriate replacement API (from \textit{ResourcesServiceImpl.verifyFile} to \textit{ResourcesService.updateResource}); Diffploit is only able to perform part of the migration, leaving the exploit nonfunctional.

\subsection{Threats to Validity}

Our study mainly suffers from the following threats to validity:

\textbf{External Validity} The major threat lies in the generality of the dataset. To overcome this, we evaluate \appname using the largest publicly available Java exploit dataset.
However, we acknowledge that the performance on other programming languages remains unverified. Another threat is that the affected version range is manually labeled, which may introduce human errors. Although we perform a second-round validation and eliminate 22.7\% of incorrect labeling, there may still be misclassified vulnerable versions.
\appname relies on source code to obtain diffs between library versions through the corresponding GitHub repository. However, for some projects, version diffs are no longer publicly available, such as versions of \textit{spring-framework} prior to 3.0.0. In such cases, \appname may fail due to the lack of accessible diffs.

\textbf{Internal Validity.} One potential threat to internal validity is that the performance of \appname may result from the base model’s prior exposure to migrated exploits, rather than its true generalization capability. To examine this, we conduct an experiment where only the base model is used to perform exploit migration. We observe that the success rate drops significantly when the full method is replaced with the base model alone. This result suggests that the effectiveness of \appname does not stem from potential data leakage but from its diff-based design.

Another threat to internal validity lies in \appname depends on the mvn command for building (`\textit{mvn compile}') and managing library versions (`\textit{mvn versions:use-dep-version}'), limiting its applicability to other programming languages. Further modifications to the compilation and exploitation process are required. \appname relies on `\textit{git diff}' to generate the diff file between the target and reference versions, 
which limits its applicability to other version control systems.

\section{Related Work}

Our work focuses on the task of \textbf{exploit migration}, which involves adapting existing exploits to different software versions. This task is crucial for accurate vulnerability assessment and exploit reproduction in real-world scenarios, where environmental differences and software evolution often break naive exploits. Recent studies have proposed various approaches to tackle exploit migration and cross-version exploitability assessment: AEM~\cite{Jiang2023AEM} introduces an automated exploit migration method targeting Linux kernels, which aligns execution points in different kernel versions to reproduce exploitation behaviors. Similarly, VulScope~\cite{Dai2021Exploit} leverages directed fuzzing to migrate exploits between software versions, improving vulnerability detection coverage. SyzBridge~\cite{zou2024SyzBridge} addresses environmental differences between upstream Linux kernels and downstream distributions by adapting exploits accordingly, enhancing exploitability assessment accuracy. Evocatio~\cite{Jiang2022Expand} automatically generates exploits to expose previously unknown bug capabilities. These efforts underscore the complexity and practical importance of exploit migration. The aforementioned approaches rely on fuzzing and require explicit execution trace mapping to support exploit migration, which is time-consuming and challenging.

Despite its importance in vulnerability analysis, exploit migration remains underexplored in Java vulnerabilities, motivating our approach, which handles both triggering condition change and environment breakage in a diff-driven framework.

We observe that exploit migration shares similarities with \textbf{test migration} and \textbf{API migration}. Like automated test repair~\cite{Daniel2011Test, Rahman2024flaky, Alshahwan2024meta}, automated test migration has attracted increasing attention as software systems evolve rapidly, requiring frequent updates to maintain test suite reliability and effectiveness~\cite{Daniel2009Test, Mirzaaghaei2014Repair}. Prior work includes \testrepairname, which leverages pre-trained language models to treat test repair as a language translation task~\cite{Saboor2025repair, Hoffman2009matching}. UTFix focused on repairing python unit tests affected by changes in focal methods, using contextual static and dynamic code slices and failure messages~\cite{Rahman2025UTFix}. Earlier frameworks such as TestCareAssistant repaired or generated tests by adapting to limited changes like method parameter additions ~\cite{Mirzaaghaei2011repair}, while TestFix uses genetic algorithms to synthesize method call insertions and deletions to fix broken JUnit tests, albeit only supporting single-assertion tests ~\cite{XU2014repair}. To better preserve test intent, TRIP employs a search-based approach guided by dynamic symbolic execution to prioritize repair candidates that maintain original test semantics, generating fixes for tested cases ~\cite{Li2019repair}. Similar to test migration, API migration aims at updating API usages in downstream projects to resolve library updates~\cite{Fazzini2019API, Zhong2024API, Xu2019API}. These approaches can partially address exploit failures caused by API changes during library evolution. Despite these advances, test migration and API migration remain challenging due to the diversity of exploit ineffective reasons. Moreover, existing tasks typically focus only on function-level test repair, overlooking the challenges posed by environmental factors. Addressing such environmental challenges remains an underexplored area.

\section{Conclusion}

In this work, we present \appname, a novel framework for exploit migration that combines LLM-driven adaptation with dynamic context construction based on version-specific code diffs. By identifying both causing and supporting diffs and employing a simulated annealing strategy to guide iterative adaptation, \appname transforms failure symptoms into actionable migration steps. This feedback-driven process enables reliable exploit transfer across complex version gaps. Through extensive experiments on the largest Java vulnerability exploit dataset to date, \appname achieves a high success rate of 84.2\%, outperforming state-of-the-art baselines by a significant margin. Our ablation study confirms that each design component contributes meaningfully to both effectiveness and cost-efficiency. Moreover, real-world responses from CNAs and open-source maintainers affirm the practical feasibility of our approach, with 82 affected versions accepted into GitHub Advisory Database and CNAs of three CVEs confirm affected versions identified based on our migrated exploits. While \appname demonstrates strong adaptability and generality, it remains limited in handling modified reproduction behaviors. These challenges point to promising future directions. Overall, our study shows that \appname not only enhances exploit usability across versions but also contributes to the completeness and accuracy of vulnerability databases.

\section*{Acknowledgement}
This research is supported by the National Key R\&D Program of China (No.2024YFB4506400). We also thank the anonymous reviewers for their insightful comments and suggestions.

\balance
\bibliographystyle{ACM-Reference-Format}
\bibliography{main}

\end{document}